\documentclass[conference]{IEEEtran}
\IEEEoverridecommandlockouts

\usepackage{cite}
\usepackage{amsmath,amssymb,amsfonts}
\usepackage{algorithmic}
\usepackage{graphicx}
\usepackage{textcomp}
\usepackage{xcolor}

\usepackage{physics}
\usepackage{qcircuit}

\usepackage{cleveref}
\crefname{appsec}{Appendix}{Appendices}

\graphicspath{{./figs/}}

\newcommand{\tfd}{\psi_{\vec{\alpha}, \vec{\gamma}}}

\begin{document}

\title{Engineering the Cost Function of a Variational Quantum Algorithm for Implementation on Near-Term Devices}

\author{\IEEEauthorblockN{Shavindra P. Premaratne}
\IEEEauthorblockA{\textit{Intel Labs} \\
\textit{Intel Corporation}\\
Hillsboro, OR, USA}
\and
\IEEEauthorblockN{A. Y. Matsuura}
\IEEEauthorblockA{\textit{Intel Labs} \\
\textit{Intel Corporation}\\
Hillsboro, OR, USA}
}

\maketitle

\begin{abstract}
Variational hybrid quantum-classical algorithms are some of the most promising workloads for near-term quantum computers without error correction. The aim of these variational algorithms is to guide the quantum system to a target state that minimizes a cost function, by varying certain parameters in a quantum circuit. This paper proposes a new approach for engineering cost functions to improve the performance of a certain class of these variational algorithms on today’s small qubit systems. We apply this approach to a variational algorithm that generates thermofield double states of the transverse field Ising model, which are relevant when studying phase transitions in condensed matter systems. We discuss the benefits and drawbacks of various cost functions, apply our new engineering approach, and show that it yields good agreement across the full temperature range.
\end{abstract}


\section{Introduction}

Noisy intermediate scale quantum (NISQ) computers are near-term systems that have up to a few hundred non-error-corrected qubits that suffer from decoherence and noise, limiting the overall quantum circuit depth \cite{Preskill2018}. Some of the main limitations of NISQ systems include limited qubit number, limited qubit connectivity, and hardware-specific quantum gate alphabets \cite{Khatri2019}. Algorithms must be optimized carefully to smaller quantum circuit depths in order to run on NISQ systems \cite{Khatri2019, Zou2020}. Hybrid classical-quantum variational algorithms are promising implementations for NISQ systems and include the quantum approximate optimization algorithm (QAOA) \cite{Farhi2014}, and Variational Quantum Eigensolver (VQE) \cite{Peruzzo2014, Moll2018}. These variational algorithms use classical compute resources to enable the execution of more complicated algorithms on the small quantum compute resources available today. 

Implementation of a variational algorithm typically involves the minimization of a cost function in order to obtain the optimal variational parameters to generate the desired quantum state. The purpose of the variational algorithm described in this work is to construct a quantum state called a thermofield double (TFD) state in the transverse field Ising model (TFIM), which is important in understanding thermal phase transitions in condensed matter systems. TFD states are entangled pure states between two systems which yield a thermal state when one of the systems is traced out \cite{Nielsen2010, Maziero2017}. We develop an engineering approach to construct an optimal cost function for implementation on a small qubit system, and show that cost functions engineered using our approach can result in more accurate outcomes on real hardware systems. This higher accuracy will yield better experimental results in state-of-the-art NISQ systems, as evidenced by the implementation of a simpler variant of a similarly engineered cost function to generate TFD states in an actual superconducting quantum processor \cite{Sagastizabal2020, Sagastizabal2020b}.

\section{Thermofield Double States Generation \label{sec:tfd_generation}}

\subsection{System Definition}

The Hamiltonian of the TFIM for a one-dimensional ring of $N$ qubits is given by
\begin{align}
\mathcal{H}_\textrm{TFIM} &= \sum_{i=1}^N \textsf{Z}_i \textsf{Z}_{i+1} + g \sum_{i=1}^N \textsf{X}_i = \mathcal{H}_\textsf{ZZ} + g \mathcal{H}_\textsf{X}
\label{eq:H_TFIM}
\end{align}
where the transverse field direction was chosen for convenience of implementation in superconducting systems (compare to \cite{Zhu2019} with $\textsf{X} \leftrightarrow \textsf{Z}$). We use natural units (\textit{i.e.} $\hbar = k_b = 1$) throughout the manuscript. Consider the special case of generating Thermofield Double (TFD) states in a four-qubit system which was recently demonstrated experimentally, using superconducting qubits by \cite{Sagastizabal2020}. In this case, the intra-system Hamiltonian reduces to 
\begin{align}
\mathcal{H}_\textrm{intra} &= \textsf{Z}_1 \textsf{Z}_2 + g (\textsf{X}_1 + \textsf{X}_2)
\label{eq:H_intra}
\end{align}
which describes interactions within each of the two subsystems ($A$ and $B$) with $g$ being the transverse field strength. The ultimate objective is to have the full system undergo unitary evolution such that it will yield a thermal state (or Gibbs state) on subsystem $A$, if it is considered in isolation. In practice, this can be studied by performing a partial trace over subsystem $B$ \cite{Nielsen2010}. Conversely, this technique can be viewed as a purification of the Gibbs state, resulting in a TFD in the full system \cite{Wu2019}. The TFD state $\ket{\xi}$ at an inverse temperature $\beta=T^{-1}$ is thus defined as 
\begin{align}
\ket{\xi (\beta)} \equiv \frac{1}{\sqrt{\mathcal{N}}} \exp \left(- \frac{\beta}{2} \mathcal{H}_A \right) \ket{\xi(0)}
\label{eq:tfd_definition}
\end{align}
where $\mathcal{N}$ is a normalization factor, and
\begin{align}
\mathcal{H}_A = \textsf{ZZ}_{A} + g \textsf{X}_{A} = \textsf{Z}_{A1} \textsf{Z}_{A2} + g (\textsf{X}_{A1} + \textsf{X}_{A2}).
\end{align} 
$\ket{\xi(0)}$ is the TFD state at $\beta=0$ or $T \rightarrow \infty$, which should be a pairwise maximally entangled Bell state (since tracing subsystem $B$ out of this full state yields a maximally-mixed state for subsystem $A$). For simulation purposes we thus set the initial state to be 
\begin{align}
\ket{\xi{(0)}} &= \textsf{CNOT}_{24} \cdot \textsf{CNOT}_{13} \cdot \textsf{H}_2 \cdot \textsf{H}_1 \ket{0000} \nonumber \\
&= \frac{1}{2} \left( \ket{0000} + \ket{0101} + \ket{1010} + \ket{1111} \right)
\label{eq:init_state}
\end{align}
where the qubits are labeled as $\ket{ A_1, A_2, B_1, B_2}$ and qubits $\left\{ A_i, B_i \right\}$ are pairwise maximally entangled through applied unitaries.

\subsection{System Evolution}


The protocol for generation of TFD states is described in \cite{Zhu2019}, and we follow a similar path to invoke the variational ansatz motivated by the quantum alternating operator ansatz \cite{Hadfield2019}. This involves alternating between the subsystem Hamiltonian $\mathcal{H}_A + \mathcal{H}_B$ and the entangling Hamiltonian $\mathcal{H}_{AB}$. Here the subsystem $B$ Hamiltonian is defined as
\begin{align}
\mathcal{H}_B = \textsf{ZZ}_{B} + g \textsf{X}_{B} = \textsf{Z}_{B1} \textsf{Z}_{B2} + g (\textsf{X}_{B1} + \textsf{X}_{B2})
\end{align}

\noindent and the entangling Hamiltonian is defined as
\begin{align}
\mathcal{H}_\mathit{AB} &= \textsf{XX}_{AB} + \textsf{ZZ}_{AB} \nonumber \\
&= \textsf{XX}_{AB1} + \textsf{XX}_{AB2} + \textsf{ZZ}_{AB1} + \textsf{ZZ}_{AB2} \nonumber \\
&= \textsf{X}_{A1} \textsf{X}_{B1} + \textsf{X}_{A2} \textsf{X}_{B2} + \textsf{Z}_{A1} \textsf{Z}_{B1} + \textsf{Z}_{A2} \textsf{Z}_{B2}.
\label{eq:H_inter}
\end{align}

Near-term quantum computing systems have stringent coherent operations limits and it is desirable to minimize the number of steps required for a given workload \cite{Zou2020}. Hence, here we focus on the case of increasing efficiency of single-step TFD generation under the given Hamiltonians (\cref{eq:H_intra,eq:H_inter}). With our restricted model, the quantum circuit for TFD generation is described by \cref{eq:evolution} with $\left\{ \alpha_1, \alpha_2, \gamma_1, \gamma_2 \right\}$ as variational parameters.

\small
\begin{align}
\begin{split}
\ket{\tfd} 
&= \ket{\psi \left(\alpha_1, \alpha_2, \gamma_1, \gamma_2 \right)} \\
&= e^{i \alpha_2 \textsf{ZZ}_{AB}} e^{i \alpha_1 \textsf{XX}_{AB}} e^{i \gamma_2 \left(\textsf{ZZ}_A + \textsf{ZZ}_B \right) } e^{i \gamma_1 \left(\textsf{X}_A + \textsf{X}_B \right)} \ket{\xi{(0)}}
\label{eq:evolution}
\end{split}
\end{align}
\normalsize

\section{Comparison of Cost Functions for TFD States' Generation}

To obtain the optimal variational parameters $\left\{ \alpha_1, \alpha_2, \gamma_1, \gamma_2 \right\}$ experimentally, a quantum computer is used for system evolution while a classical computer is used for cost function evaluation based on the measurements returned by the quantum computer. The ultimate success of the protocol depends on the effectiveness of the cost function evaluation as well as the capability of the quantum circuit to accurately generate the desired TFD state. 

\subsection{Error in fidelity between generated and target states}

For numerical convenience it is typical to utilize the error in fidelity $\mathcal{E}$ with respect to the target as the cost function during optimization \cite{Wu2019},
\begin{align}
\mathcal{E} &= 1- \left| \bra{\xi(\beta)} \ket{\tfd} \right|^2.
\label{eq:overlap_fidelity}
\end{align}
Evaluation of the latter expression requires access to the actual wavefunction of the generated state. Experimental determination of the wavefunction is typically limited to either estimation based on extensive tomographic reconstruction \cite{Altepeter2004}, or an array of sophisticated weak measurements \cite{Lundeen2011}. Thus it is undesirable in its present form for evaluation in an actual hybrid quantum-classical experimental system. In principle, it is possible to modify \cref{eq:overlap_fidelity} to use the density matrix of the generated state, but this also is cumbersome and reasons are discussed further in the next section.

\subsection{Free energy of the system \label{ssec:f_a}}

An alternative to \cref{eq:overlap_fidelity} is using the free energy of the system as the cost function during optimization \cite{Wu2019}. In this case, the free energy $F_A$ is calculated on the reduced density matrix for subsystem $A$ as
\begin{align}
F_A (T; \vec{\alpha}, \vec{\gamma}) &= E_A - T S_A \nonumber \\
&= \Tr \left[ \varsigma_A H_A \right] + T \Tr \left[ \varsigma_A \log \varsigma_A \right],
\label{eq:free_energy}
\end{align}
where $E_A$, $S_A$, $\varsigma_A = \Tr_B \ketbra{\tfd}$ are the energy, von Neumann entropy, and the reduced density operator for subsystem $A$, respectively, and $T$ is the system temperature. In this case quantum state tomography (QST) \cite{Altepeter2004} of subsystem $A$ is necessary to calculate $F_A$. Recently, methods to approximate the von Neumann entropy have also been proposed \cite{Chowdhury2020}.

Typically, QST of a system requires a complete set of measurements related to the number of unknowns of the system size \cite{James2001}. Given a system of $n_\textrm{q}$ qubits, the number of unique density matrix elements is given by $\left(2^{2n_\textrm{q}} - 1 \right)$. Thus, this is the total number of unique measurements required to fully characterize the qubit system (\textit{e.g.} for a two-qubit subsystem, this gives 15 measurements). As system size increases, QST becomes prohibitively expensive and renders free energy $F_A$ a poor choice for large-scale optimization problems.

\subsection{Hypothesized cost function based on correlator expectation values}

To formulate a more experiment-friendly expression, it is possible to hypothesize a cost function based on the form of $F_A$ as described in \cref{ssec:f_a}. This would involve substituting alternative expressions for $E_A$ and $S_A$ that require fewer measurements than QST. It is straightforward to infer that $F_A$ will be dominated by $S_A$ at higher temperatures, and by $E_A$ at lower temperatures. Given the simplicity of the TFIM Hamiltonian, the expression for the low temperature regime is easily obtained as follows,
\begin{align}
E_A &= \expval{\mathcal{H}_A} = \expval{\textsf{Z}_{A1} \textsf{Z}_{A2} + g (\textsf{X}_{A1} + \textsf{X}_{A2})} \nonumber \\
&= \expval{\textsf{Z}_{A1} \textsf{Z}_{A2}} + g \expval{\textsf{X}_{A1}} + g \expval{\textsf{X}_{A2}}
\label{eq:E_A}
\end{align}
where the $\expval{\cdot}$ notation indicates the expectation value of the relevant correlator with respect to the evolved wavefunction $\ket{\tfd}$ in \cref{eq:evolution}. The reduction in the number of terms is not strictly due to the use of correlators, but it is rather the underlying symmetry of the problem that allows us to determine $E_A$ with only three system measurements for subsystem $A$. However, this method of constructing the cost function allows an explicit represention using tangible measurements for the hybrid quantum-classical optimization algorithm. 

There is no straightforward method to derive an expression for $S_A$ based on correlator expectation values as before. However, it is trivial to verify that $\ket{\xi(0)}$ is the ground state of the negative of the inter-system Hamiltonian, $-\mathcal{H}_{AB}$. Given that we expect $S_A$ to dominate at $T\rightarrow \infty$ and $\ket{\xi(0)}$ is the infinite temperature TFD state, it is natural to hypothesize an approximate form of $S_A$ to be generalized to $\mathcal{H}_{AB}$ when considering the full system (\textit{i.e.} both $A$ and $B$) \cite{Premaratne2020}. This results in the following expression for a cost function $\mathcal{C}_0 (T)$, which is more amenable for practical implementation in a quantum-classical optimization algorithm.
\begin{align}
\mathcal{C}_0 (T; \vec{\alpha}, \vec{\gamma}) &= \mel**{\tfd}{\mathcal{H}_A + \mathcal{H}_B - T \mathcal{H}_{AB}}{\tfd},
\label{eq:C_0}
\end{align}
where the system temperature $T$ is a parameter, and $\left\{ \alpha_1, \alpha_2, \gamma_1, \gamma_2 \right\}$ are optimization variables. Here, we have generalized the cost function to the full system and included $\mathcal{H}_B$ in the energy of the system at low temperatures.

\subsection{Free Energy vs. Hypothesized Cost Function \label{ssec:FA_vs_C0}}

When the strength of the transverse field is set to $g=1$, a critical point between antiferromagnetic and paramagnetic quantum phases is expected \cite{Bonfim2019, Pfeuty1970}. Hence, for cost function performance comparison purposes, we will primarily explore the case of $g=1$. The case of $g \neq 1$ will be considered for completeness in \cref{app:g_neq_1}. We simulate TFD state generation using Differential Evolution, which is a global optimization algorithm \cite{Storn1997} supported by Wolfram Mathematica for non-linear optimization. The optimization is performed over a wide inverse temperature range of six orders of magnitude to ensure complete coverage. 

\begin{figure}[tb!]
\includegraphics[width=\columnwidth]{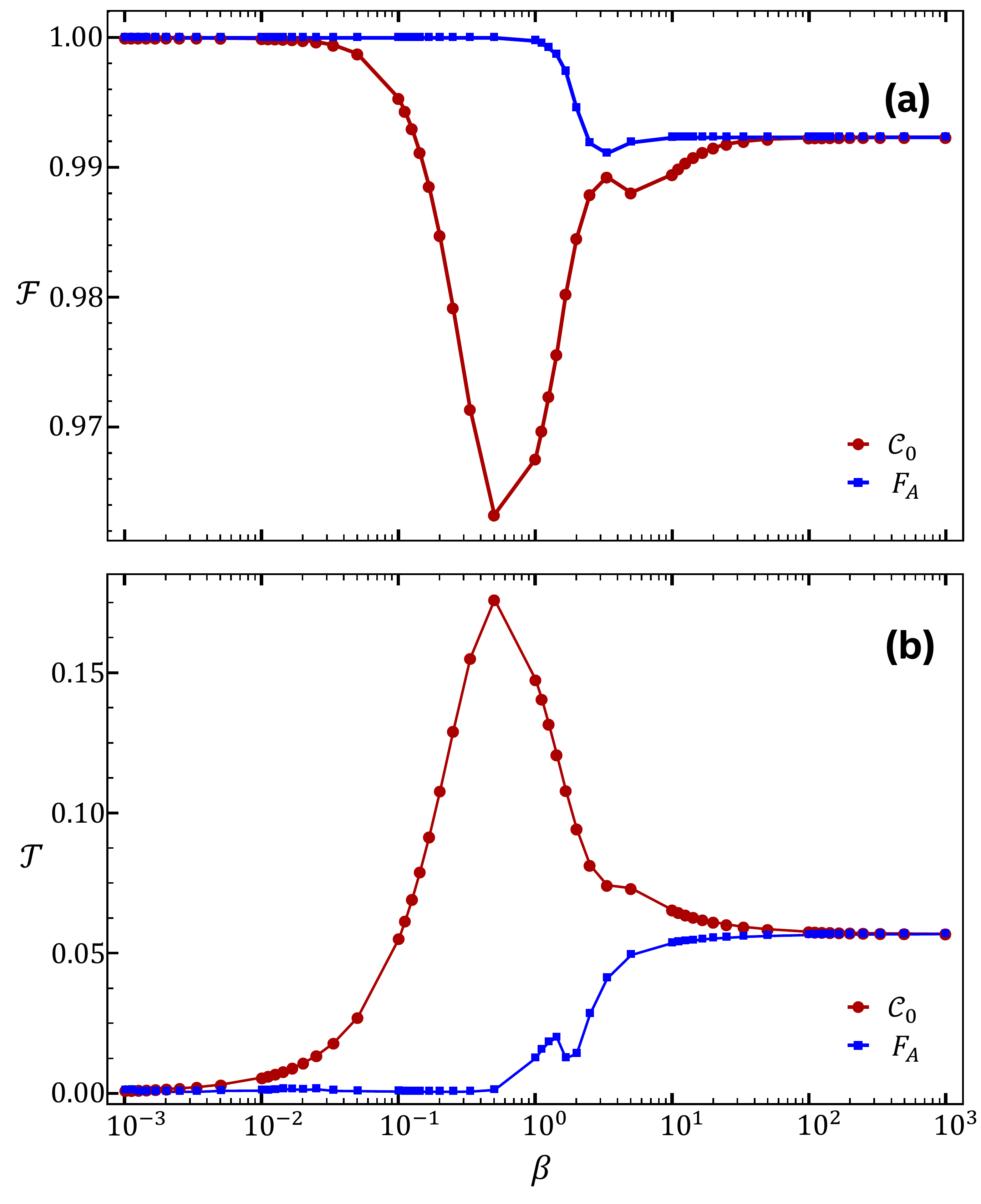}
\caption{Cost function performance comparison between $F_A$ and $\mathcal{C}_0$, using (a) fidelity $\mathcal{F}$ and (b) trace distance $\mathcal{T}$ as proximity measures.}
\label{fig:C_0-vs-F_A}
\end{figure}

During optimization we minimize the cost functions corresponding to $F_A$ and $\mathcal{C}_0$ separately, and see that there is excellent agreement between them for extreme low and high temperatures (see \cref{fig:C_0-vs-F_A}). We have chosen trace distance $\mathcal{T}$ and fidelity $\mathcal{F}$ as defined below \cite{Nielsen2010}, as proximity measures comparing the ideal TFD state and the optimally generated state with the different cost functions.
\begin{align}
\mathcal{T} &= \frac{1}{2} \Tr \left[ \sqrt{(\rho_A - \sigma_A)^{\dagger} \cdot (\rho_A - \sigma_A)} \right] \\
\mathcal{F} &= \left( \Tr \sqrt{\sqrt{\rho_A} \sigma_A \sqrt{\rho_A}} \right)^2
\end{align}
Here, $\rho = \ketbra{\xi(\beta)}$ is the density matrix corresponding to the ideal TFD state, $\sigma = \ketbra{\Psi(\beta)}$ represents the density matrix for the circuit-generated state $\ket{\Psi(\beta)}$ utilizing the relevant cost function, and $\rho_A / \sigma_A$ are corresponding subsystem states after tracing out $B$, respectively.

For perfect TFD state preparation, we would expect $\mathcal{T} = 0$ and $\mathcal{F} = 1$. However, in the range $10^{-2} < \beta < 10^{2}$, the performance of $\mathcal{C}_0$ as a cost function is poor, and demonstrates difficulty in reaching the target TFD state. This is somewhat expected, given that we hypothesized the form of the simple cost function based on extreme high/low temperature behavior. Also note that for $\beta \rightarrow \infty$, $\mathcal{F} \neq 1$ and $\mathcal{T} \neq 0$ indicating that even while using free energy as the cost function we do not construct the ideal TFD state. This is also expected since $T \rightarrow 0$ is the most difficult regime to generate the TFD state based on the given protocol \cite{Wu2019}. This indicates that the depth of the circuit is most likely insufficient to construct a better state approaching the TFD state. However depending on the measure used, we find that better approximations to the TFD states are possible using different engineered cost functions.

\section{Engineering Improved Cost Functions}

\subsection{Enhancing the hypothesized cost function \label{ssec:mod_corr}}

\begin{figure}[tb!]
\includegraphics[width=\columnwidth]{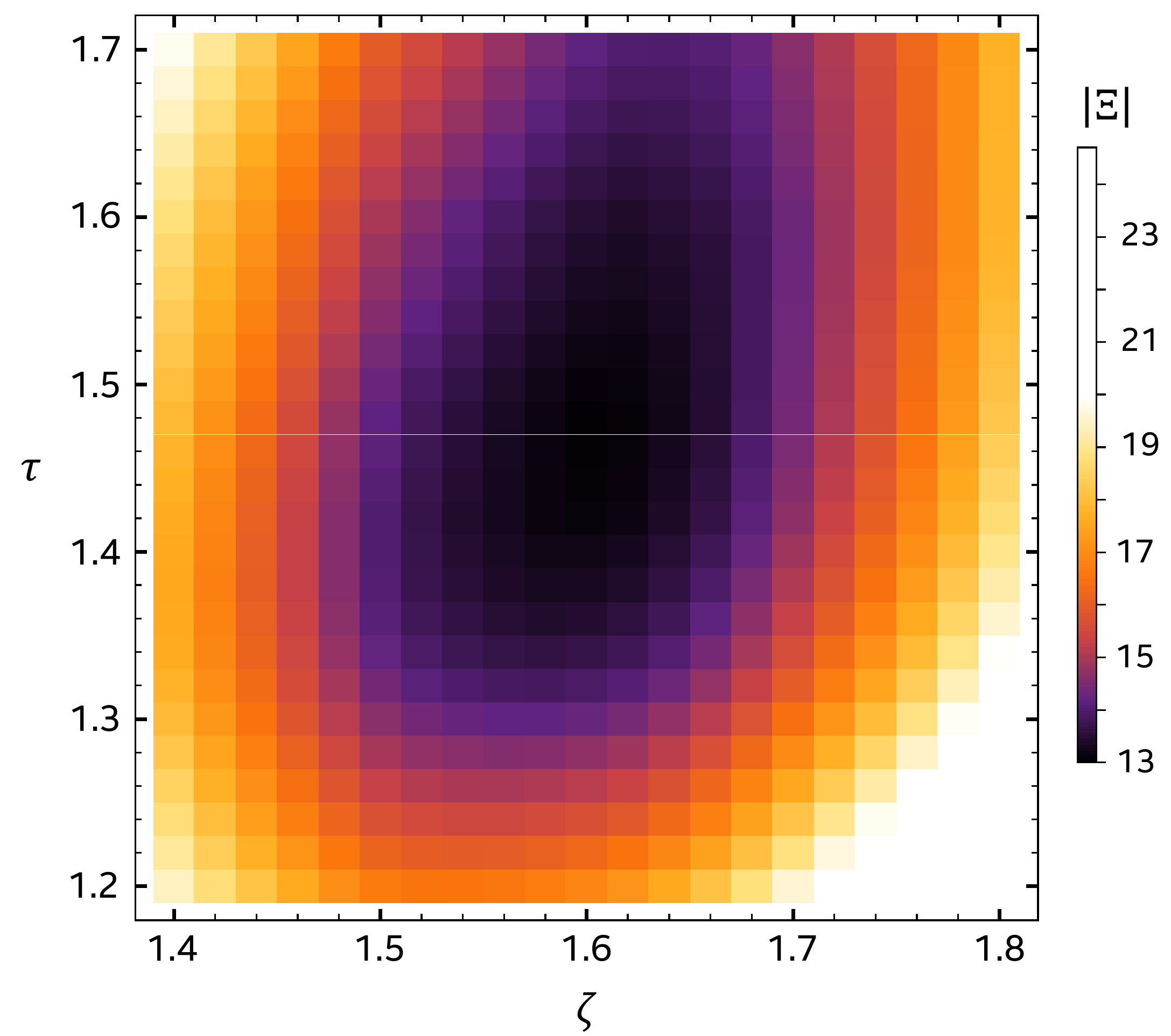}
\caption{Two-dimensional sweep plot of $\left| \Xi \right|$ vs. $\zeta$ and $\tau$ to enhance the hypothesized cost function. The minimum value for $\left|\Xi \right|$ is observed for $\zeta=1.6$ and $\tau = 1.48$.}
\label{fig:mod_corr}
\end{figure}

Given the shortcomings of $\mathcal{C}_0$ at intermediate $\beta$ values, we began our new approach to construct a better cost function by generalizing \cref{eq:C_0}  for $g=1$ by adding coefficients to the expression containing correlators as follows,
\begin{align}
\mathcal{C}_1(T; \vec{\alpha}, \vec{\gamma}) = \mel**{\tfd}{\textsf{c}_1}{\tfd} ,
\end{align}
where 
\begin{align}
\textsf{c}_1 (\zeta, \tau) &= \textsf{X}_{A} + \textsf{X}_{B} + \zeta \left( \textsf{ZZ}_{A} + \textsf{ZZ}_{B} \right) \nonumber \\ 
&\hspace{7em} - T^\tau \left(\textsf{ZZ}_{AB} + \textsf{XX}_{AB}\right).
\end{align}
Here $\zeta$ and $\tau$ are parameters which we optimize to find a cost function $\mathcal{C}_1$ that can yield better approximations to TFD states across the full inverse temperature range. For simplicity and clarity, instead of performing nested optimizations over $\left\{ \zeta, \tau \right\}$ and $\left\{\vec{\alpha}, \vec{\gamma} \right \}$, we execute TFD generation for a range of $\zeta$ and $\tau$ values (see \cref{app:minimization_quantity} for details). By varying $1.4 < \zeta < 1.9$, and $1.2 < \tau < 1.7$, and evaluating the minimization quantity of interest $\Xi$, we see that the best agreement between $\ket{\xi(\beta)}$ and $\ket{\Psi(\beta)}$ is obtained for $\zeta=1.6$ and $\tau = 1.48$ (see \cref{fig:mod_corr}). The improvements from using $\mathcal{C}_1$ are discussed in \cref{ssec:cf_comparison}.

\subsection{Analytically obtaining a better cost function \label{ssec:pruned_elements}}

To further the work towards engineering a better cost function, we used an alternative approach based on density matrix elements' closeness to generate a cost function that shows significant improvement over both the hypothesized cost function $\mathcal{C}_0$, and its enhanced version $\mathcal{C}_1$. In this case, we studied the characteristics of the density matrices of the target TFD state, $\rho$, and the single-step circuit-generated state, $\sigma$. Note that we are not tracing out subsystem $B$ of these matrices to obtain the thermal state. Instead we directly compare the density matrices of the target and generated states. Also we relax the assumption of $g=1$, and keep $g$ as a parameter throughout the analysis.

First, we observe the redundancies present in the ideal TFD state and obtain 15 unique real elements for $\rho$:
\begin{align}
\begin{split}
\mathcal{R} &= \{ \rho_{00}, \rho_{01}, \rho_{03}, \rho_{05}, \rho_{06}, \rho_{11}, \rho_{13}, \rho_{15}, \rho_{16}, \\
&\hspace{2em} \rho_{33}, \rho_{35}, \rho_{36}, \rho_{55}, \rho_{56}, \rho_{66} \}
\end{split}
\end{align}
where the density matrix elements are labeled consistent with the notation in \cref{eq:init_state} (see \cref{app:dm} for details). Similarly, we observe the redundancies and Hermiticity in $\sigma$ to obtain 10 unique complex elements for off-diagonal elements, and 5 unique real elements for the diagonal elements:
\begin{align}
\begin{split}
\mathcal{S} &= \{ \sigma_{00}, \sigma_{01}, \sigma_{03}, \sigma_{05}, \sigma_{06}, \sigma_{11}, \sigma_{13}, \sigma_{15}, \sigma_{16}, \\
&\hspace{2em} \sigma_{33}, \sigma_{35}, \sigma_{36}, \sigma_{55}, \sigma_{56}, \sigma_{66} \}
\end{split}
\end{align}
Inspired by trial optimizations, we find that $\sigma$ is explicitly symmetrized by choosing particular values for the inter-system variational parameters $\vec{\alpha}$,
\begin{align}
\begin{split}
\alpha_1 &= \pi/8 \\
\alpha_2 &= \pi/4
\end{split}
\end{align}
resulting in 14 unique real elements in $\sigma$. With this choice for $\vec{\alpha}$, it is found that $\sigma_{35} = \sigma_{06}$, indicating that this constrained $\sigma$ will be limited in fully matching $\rho$.

Following symmetrization, a cost function is constructed explicitly by calculating the sum of the square of differences between the density matrix elements for $\rho$ and $\sigma$:
\begin{align}
\mathcal{C}_2 = \sum_{i=1}^{15} a_i (r_i - s_i)^2 \label{eq:pruned_cf}
\end{align}
where $r_i \in \mathcal{R}$, $s_i \in \mathcal{S}$, and $a_i \in \{0,1 \}$ is a weight used to prune elements. We find that choosing 
\begin{align}
a_i = \left\{ 
\begin{array}{cl}
1 , & i \in \{4, 8, 13\} \\
0 , & i \in \{1, 2, 3, 5, 6, 7, 9, 10, 11, 12, 14, 15\}
\end{array}
\right.
\end{align}
generates a simple cost function which yields good results across the full temperature range and for different $g$ values. For this system of four qubits, we find that it is beneficial to substitute variational angles in the engineered cost function $\mathcal{C}_2$ with the operator expectation values as was the case in $\mathcal{C}_0$ and $\mathcal{C}_1$. This will enable the experiments to be performed using identical quantum processor measurements, while modifying the classical processor evaluation to improve efficiency. Thus, the non-zero elements for the engineered cost function $\mathcal{C}_2$ are explicitly given by \cref{eq:pruned_elements}. Note that given the symmetric nature of the evolution of the subsystems, it is sufficient to include only the intra-system expectation values for one subsystem.

\subsection{Improvements from Engineered Cost Functions\label{ssec:cf_comparison}}

\begin{figure}[tb!]
\includegraphics[width=\columnwidth]{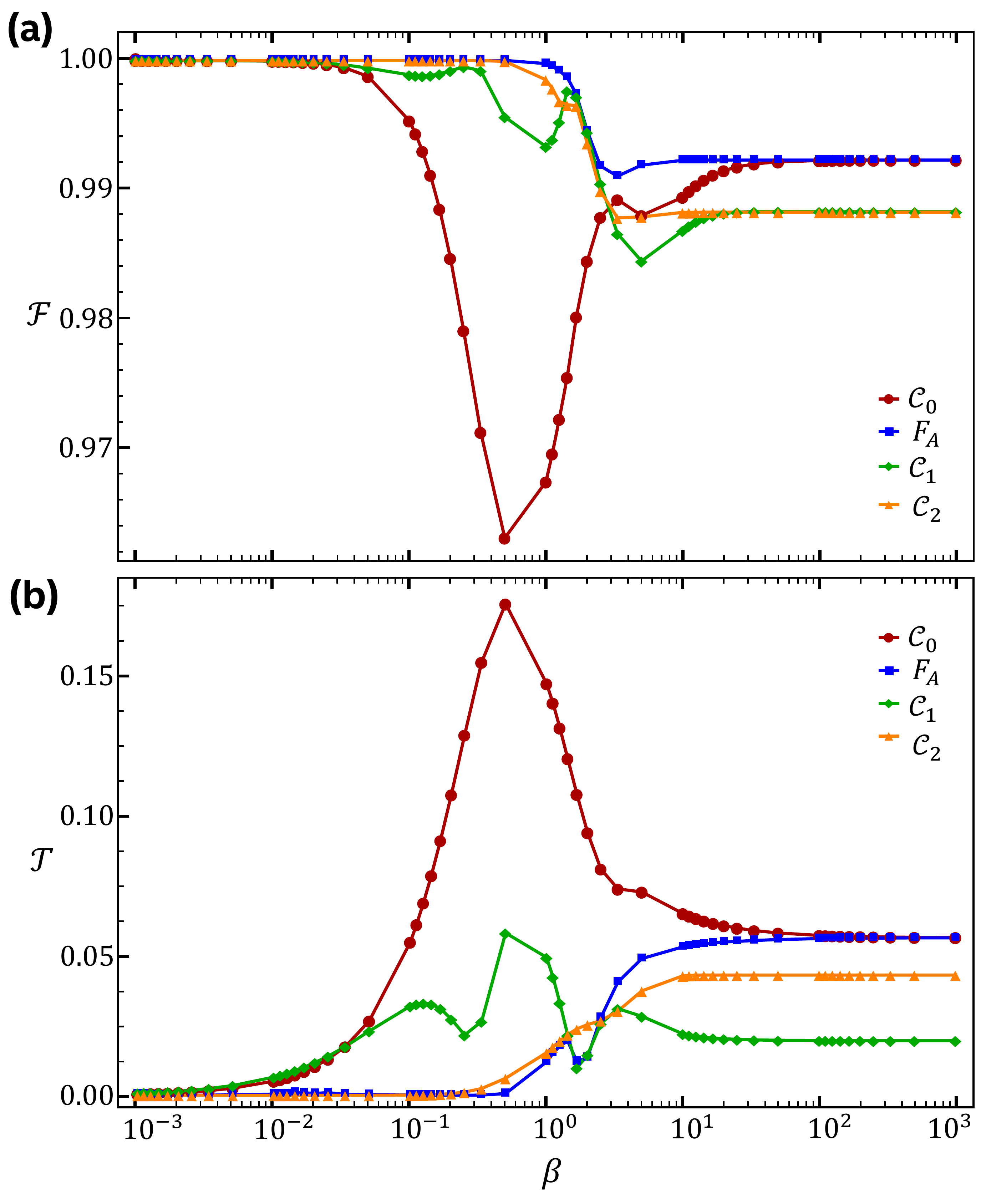}
\caption{Cost function performance comparison between $F_A$, $\mathcal{C}_0$, $\mathcal{C}_1$, and $\mathcal{C}_2$ using (a) fidelity $\mathcal{F}$ and (b) trace distance $\mathcal{T}$ as proximity measures.}
\label{fig:All_CF}
\end{figure}

We now evaluate the performance of the various cost functions when generating the TFD states for a wide temperature range. Fidelity and trace distances are calculated for the traced out subsystem as described in \cref{ssec:FA_vs_C0}. In \cref{fig:All_CF}, we observe that $\mathcal{C}_1$ yields vastly superior results compared to the original cost function $\mathcal{C}_0$, especially at intermediate temperatures. We also find that $\mathcal{C}_2$ performs significantly better than $\mathcal{C}_1$ at high temperatures. 

For $\beta > 1$, we note that the two measures ($\mathcal{F}$ and $\mathcal{T}$) offer somewhat inconsistent results. When considering trace distance $\mathcal{T}$ as the proximity measure, $\mathcal{C}_1$ and $\mathcal{C}_2$ seems to give better results compared to both $\mathcal{C}_0$ and $F_A$. In \cref{app:g_neq_1} we study a few cases of $g \neq 1$ and find that $\mathcal{C}_2$ outperforms all other cost functions (including $F_A$) irrespective of the proximity measure used. We attribute this anomaly to the definition of the measures themselves, and further study of this aspect is beyond the scope of this work.

\section{Conclusion}

In this article, we explored different cost functions that can be used during hybrid classical-quantum variational algorithm execution for TFD generation on real qubit systems. NISQ systems are constrained due to imperfect quantum operations with relatively low fidelities, low qubit lifetimes, and small qubit numbers. Hence, it is necessary to implement quantum algorithms in the most effective manner to utilize the available resources to their fullest extent. Our aim was to engineer a cost function which will generate the TFD states in a NISQ four-qubit system, in the most efficient manner. The constructed cost functions yielded a substantial improvement over the original cost function results (\textit{e.g.} $>80\%$ reduction in relative error for intermediate temperature TFD states).

The originally hypothesized cost function $\mathcal{C}_0$ was found to be inadequate in approximating the TFD state at intermediate temperatures. The enhanced cost function $\mathcal{C}_1$ obtained via modifications to the original cost function yielded better results at intermediate temperatures for the quantum critical case of $g=1$. Subsequently, the cost function $\mathcal{C}_2$ constructed based on the closeness of density matrix elements yielded good results for the full temperature range and for the transverse field range $g \in \{-0.1, -0.2, -0.5, 1, 2, 5\}$.

The method of improving the cost function to formulate $\mathcal{C}_1$ as discussed in \cref{ssec:mod_corr} is amenable to experimental exploration of cost function generation, and could result in novel methods for obtaining better cost functions for special state generation. This method can be further extended by adding other correlator expectation values in the cost function with coefficients to be found experimentally. As quantum state evolution typically incurs the highest resource cost during the experiment, this method of constructing cost functions could lead to more efficient scaling of variational algorithms to higher qubit numbers.

Although \cref{eq:pruned_elements} only included the intra-system expectation values for one subsystem, it is possible to include the other subsystem's measurements when performing an experiment. Judicious choice of the measurements between the different subsystems should allow higher throughput of measurements for faster variational algorithm execution. Although the construction of $\mathcal{C}_2$ is not a scalable technique for higher qubit numbers, excellent results were obtained for all temperature regimes. In $\mathcal{C}_2$ construction only 20\% of the density matrix elements evaluated for closeness, indicating that the encoding of the thermal state is primarily in a few populations and coherences of the TFD state. Studying how this encoding space will scale with qubit number should shed light on methods to improve practical TFD generation.

\appendices

\crefalias{section}{appsec}

\section{Definition of Minimization Quantity of Interest for Improving the hypothesized cost function
\label{app:minimization_quantity}}

The list of 15 operators considered in \cref{ssec:mod_corr} to compute the minimization quantity of interest is given by,
\begin{align}
\begin{split}
\mathcal{O} &= \left\{ \mathsf{X}_A, \mathsf{X}_B, \mathsf{Y}_A, \mathsf{Y}_B, \mathsf{Z}_A, \mathsf{Z}_B, \right. \\
&\hspace{2em} \mathsf{XX}_A, \mathsf{XX}_B, \mathsf{YY}_A, \mathsf{YY}_B, \mathsf{ZZ}_A, \mathsf{ZZ}_B, \\
&\hspace{2em} \left. \mathsf{XX}_{AB}, \mathsf{YY}_{AB}, \mathsf{ZZ}_{AB}\right\}
\end{split}
\label{eq:correlator_operators}
\end{align}
and the range of 55 temperatures used for the optimization is,
\begin{align}
\begin{split}
\mathcal{B} &= \{ 10^{-3} \times \{1, 2, 3, 4, 5, 6, 7, 8, 9\} , \\
&\hspace{2em} 10^{-2} \times \{1, 2, 3, 4, 5, 6, 7, 8, 9\} , \\
&\hspace{2em} 10^{-1} \times \{1, 2, 3, 4, 5, 6, 7, 8, 9\} , \\
&\hspace{2em} 10^{0} \times \{1, 2, 3, 4, 5, 6, 7, 8, 9\} , \\
&\hspace{2em} 10^{1} \times \{1, 2, 3, 4, 5, 6, 7, 8, 9\} , \\
&\hspace{2em} 10^{2} \times \{1, 2, 3, 4, 5, 6, 7, 8, 9\} , \\
&\hspace{2em} 10^3 \}.
\end{split}
\end{align}

In \cref{ssec:mod_corr}, we calculate the differences in operator expectation values between the ideal TFD state and the circuit-generated state. The total of the absolute differences (for all temperatures) between each ideal and generated state is chosen as the minimization quantity of interest, $\Xi$, for finding optimal $\zeta$ and $\tau$ values:
\begin{align}
\Xi = \sum_{\beta \in \mathcal{B}} \sum_{o_i \in \mathcal{O}} \left| \ev{o_i}{\xi(\beta)} - \ev{o_i}{\Psi (\beta)} \right|
\end{align}
where $\ket{\xi(\beta)}$ and $\ket{\Psi(\beta)}$ are the ideal and circuit-generated states, respectively for each $\beta$ value. This is an expression that is more conducive for experimental implementation as well depending on the ease of obtaining various operator expectation values.

\section{Density Matrix Analysis for TFD Generation \label{app:dm}}

We label the density matrices corresponding to the full system in the binary ordering of the ket occupation states as defined in \cref{sec:tfd_generation}. For example the relevant elements for $\rho_{00}, \rho_{55}$, and $\rho_{16}$ can be found as follows,
\begin{align*}
\rho_{00} &= \ketbra{0000} \\
\rho_{55} &= \ketbra{0101} \\
\rho_{16} &= \ketbra{0001}{0110}.
\end{align*}
In \cref{eq:rho}, each unique label in $\rho$ is assigned when first encountered dduring enumeration of the matrix elements. Note that this is a symmetric matrix as expected from the definition of the purified thermal state in \cref{eq:tfd_definition}. Conversely, the circuit-generated state $\sigma$ is given by a Hermitian matrix as seen in \cref{eq:sigma}. The non-zero elements in the evaluation of \cref{eq:pruned_cf} are given by the elements in \cref{eq:pruned_elements}.

\begin{figure*}[!t]
\small
\begin{align}
\rho = \left(
\begin{array}{cccccccccccccccc}
 \rho_{00} & \rho_{01} & \rho_{01} & \rho_{03} & \rho_{01} & \rho_{05} & \rho_{06} & \rho_{01} & \rho_{01} & \rho_{06} & \rho_{05} & \rho_{01} & \rho_{03} & \rho_{01} & \rho_{01} & \rho_{00} \\
 \rho_{01} & \rho_{11} & \rho_{11} & \rho_{13} & \rho_{11} & \rho_{15} & \rho_{16} & \rho_{11} & \rho_{11} & \rho_{16} & \rho_{15} & \rho_{11} & \rho_{13} & \rho_{11} & \rho_{11} & \rho_{01} \\
 \rho_{01} & \rho_{11} & \rho_{11} & \rho_{13} & \rho_{11} & \rho_{15} & \rho_{16} & \rho_{11} & \rho_{11} & \rho_{16} & \rho_{15} & \rho_{11} & \rho_{13} & \rho_{11} & \rho_{11} & \rho_{01} \\
 \rho_{03} & \rho_{13} & \rho_{13} & \rho_{33} & \rho_{13} & \rho_{35} & \rho_{36} & \rho_{13} & \rho_{13} & \rho_{36} & \rho_{35} & \rho_{13} & \rho_{33} & \rho_{13} & \rho_{13} & \rho_{03} \\
 \rho_{01} & \rho_{11} & \rho_{11} & \rho_{13} & \rho_{11} & \rho_{15} & \rho_{16} & \rho_{11} & \rho_{11} & \rho_{16} & \rho_{15} & \rho_{11} & \rho_{13} & \rho_{11} & \rho_{11} & \rho_{01} \\
 \rho_{05} & \rho_{15} & \rho_{15} & \rho_{35} & \rho_{15} & \rho_{55} & \rho_{56} & \rho_{15} & \rho_{15} & \rho_{56} & \rho_{55} & \rho_{15} & \rho_{35} & \rho_{15} & \rho_{15} & \rho_{05} \\
 \rho_{06} & \rho_{16} & \rho_{16} & \rho_{36} & \rho_{16} & \rho_{56} & \rho_{66} & \rho_{16} & \rho_{16} & \rho_{66} & \rho_{56} & \rho_{16} & \rho_{36} & \rho_{16} & \rho_{16} & \rho_{06} \\
 \rho_{01} & \rho_{11} & \rho_{11} & \rho_{13} & \rho_{11} & \rho_{15} & \rho_{16} & \rho_{11} & \rho_{11} & \rho_{16} & \rho_{15} & \rho_{11} & \rho_{13} & \rho_{11} & \rho_{11} & \rho_{01} \\
 \rho_{01} & \rho_{11} & \rho_{11} & \rho_{13} & \rho_{11} & \rho_{15} & \rho_{16} & \rho_{11} & \rho_{11} & \rho_{16} & \rho_{15} & \rho_{11} & \rho_{13} & \rho_{11} & \rho_{11} & \rho_{01} \\
 \rho_{06} & \rho_{16} & \rho_{16} & \rho_{36} & \rho_{16} & \rho_{56} & \rho_{66} & \rho_{16} & \rho_{16} & \rho_{66} & \rho_{56} & \rho_{16} & \rho_{36} & \rho_{16} & \rho_{16} & \rho_{06} \\
 \rho_{05} & \rho_{15} & \rho_{15} & \rho_{35} & \rho_{15} & \rho_{55} & \rho_{56} & \rho_{15} & \rho_{15} & \rho_{56} & \rho_{55} & \rho_{15} & \rho_{35} & \rho_{15} & \rho_{15} & \rho_{05} \\
 \rho_{01} & \rho_{11} & \rho_{11} & \rho_{13} & \rho_{11} & \rho_{15} & \rho_{16} & \rho_{11} & \rho_{11} & \rho_{16} & \rho_{15} & \rho_{11} & \rho_{13} & \rho_{11} & \rho_{11} & \rho_{01} \\
 \rho_{03} & \rho_{13} & \rho_{13} & \rho_{33} & \rho_{13} & \rho_{35} & \rho_{36} & \rho_{13} & \rho_{13} & \rho_{36} & \rho_{35} & \rho_{13} & \rho_{33} & \rho_{13} & \rho_{13} & \rho_{03} \\
 \rho_{01} & \rho_{11} & \rho_{11} & \rho_{13} & \rho_{11} & \rho_{15} & \rho_{16} & \rho_{11} & \rho_{11} & \rho_{16} & \rho_{15} & \rho_{11} & \rho_{13} & \rho_{11} & \rho_{11} & \rho_{01} \\
 \rho_{01} & \rho_{11} & \rho_{11} & \rho_{13} & \rho_{11} & \rho_{15} & \rho_{16} & \rho_{11} & \rho_{11} & \rho_{16} & \rho_{15} & \rho_{11} & \rho_{13} & \rho_{11} & \rho_{11} & \rho_{01} \\
 \rho_{00} & \rho_{01} & \rho_{01} & \rho_{03} & \rho_{01} & \rho_{05} & \rho_{06} & \rho_{01} & \rho_{01} & \rho_{06} & \rho_{05} & \rho_{01} & \rho_{03} & \rho_{01} & \rho_{01} & \rho_{00} \\
\end{array}
\right) \label{eq:rho}
\end{align}
\vspace*{2pt}
\normalsize
\end{figure*}

\begin{figure*}
\small
\begin{align}
\sigma = \left(
\begin{array}{cccccccccccccccc}
 \sigma_{00} & \sigma_{01} & \sigma_{01} & \sigma_{03} & \sigma_{01} & \sigma_{05} & \sigma_{06} & \sigma_{01} & \sigma_{01} & \sigma_{06} & \sigma_{05} & \sigma_{01} & \sigma_{03} & \sigma_{01} & \sigma_{01} & \sigma_{00} \\
 \sigma_{01}^* & \sigma_{11} & \sigma_{11} & \sigma_{13} & \sigma_{11} & \sigma_{15} & \sigma_{16} & \sigma_{11} & \sigma_{11} & \sigma_{16} & \sigma_{15} & \sigma_{11} & \sigma_{13} & \sigma_{11} & \sigma_{11} & \sigma_{01}^* \\
 \sigma_{01}^* & \sigma_{11} & \sigma_{11} & \sigma_{13} & \sigma_{11} & \sigma_{15} & \sigma_{16} & \sigma_{11} & \sigma_{11} & \sigma_{16} & \sigma_{15} & \sigma_{11} & \sigma_{13} & \sigma_{11} & \sigma_{11} & \sigma_{01}^* \\
 \sigma_{03}^* & \sigma_{13}^* & \sigma_{13}^* & \sigma_{33} & \sigma_{13}^* & \sigma_{35} & \sigma_{36} & \sigma_{13}^* & \sigma_{13}^* & \sigma_{36} & \sigma_{35} & \sigma_{13}^* & \sigma_{33} & \sigma_{13}^* & \sigma_{13}^* & \sigma_{03}^* \\
 \sigma_{01}^* & \sigma_{11} & \sigma_{11} & \sigma_{13} & \sigma_{11} & \sigma_{15} & \sigma_{16} & \sigma_{11} & \sigma_{11} & \sigma_{16} & \sigma_{15} & \sigma_{11} & \sigma_{13} & \sigma_{11} & \sigma_{11} & \sigma_{01}^* \\
 \sigma_{05}^* & \sigma_{15}^* & \sigma_{15}^* & \sigma_{35}^* & \sigma_{15}^* & \sigma_{55} & \sigma_{56} & \sigma_{15}^* & \sigma_{15}^* & \sigma_{56} & \sigma_{55} & \sigma_{15}^* & \sigma_{35}^* & \sigma_{15}^* & \sigma_{15}^* & \sigma_{05}^* \\
 \sigma_{06}^* & \sigma_{16}^* & \sigma_{16}^* & \sigma_{36}^* & \sigma_{16}^* & \sigma_{56}^* & \sigma_{66} & \sigma_{16}^* & \sigma_{16}^* & \sigma_{66} & \sigma_{56}^* & \sigma_{16}^* & \sigma_{36}^* & \sigma_{16}^* & \sigma_{16}^* & \sigma_{06}^* \\
 \sigma_{01}^* & \sigma_{11} & \sigma_{11} & \sigma_{13} & \sigma_{11} & \sigma_{15} & \sigma_{16} & \sigma_{11} & \sigma_{11} & \sigma_{16} & \sigma_{15} & \sigma_{11} & \sigma_{13} & \sigma_{11} & \sigma_{11} & \sigma_{01}^* \\
 \sigma_{01}^* & \sigma_{11} & \sigma_{11} & \sigma_{13} & \sigma_{11} & \sigma_{15} & \sigma_{16} & \sigma_{11} & \sigma_{11} & \sigma_{16} & \sigma_{15} & \sigma_{11} & \sigma_{13} & \sigma_{11} & \sigma_{11} & \sigma_{01}^* \\
 \sigma_{06}^* & \sigma_{16}^* & \sigma_{16}^* & \sigma_{36}^* & \sigma_{16}^* & \sigma_{56}^* & \sigma_{66} & \sigma_{16}^* & \sigma_{16}^* & \sigma_{66} & \sigma_{56}^* & \sigma_{16}^* & \sigma_{36}^* & \sigma_{16}^* & \sigma_{16}^* & \sigma_{06}^* \\
 \sigma_{05}^* & \sigma_{15}^* & \sigma_{15}^* & \sigma_{35}^* & \sigma_{15}^* & \sigma_{55} & \sigma_{56} & \sigma_{15}^* & \sigma_{15}^* & \sigma_{56} & \sigma_{55} & \sigma_{15}^* & \sigma_{35}^* & \sigma_{15}^* & \sigma_{15}^* & \sigma_{05}^* \\
 \sigma_{01}^* & \sigma_{11} & \sigma_{11} & \sigma_{13} & \sigma_{11} & \sigma_{15} & \sigma_{16} & \sigma_{11} & \sigma_{11} & \sigma_{16} & \sigma_{15} & \sigma_{11} & \sigma_{13} & \sigma_{11} & \sigma_{11} & \sigma_{01}^* \\
 \sigma_{03}^* & \sigma_{13}^* & \sigma_{13}^* & \sigma_{33} & \sigma_{13}^* & \sigma_{35} & \sigma_{36} & \sigma_{13}^* & \sigma_{13}^* & \sigma_{36} & \sigma_{35} & \sigma_{13}^* & \sigma_{33} & \sigma_{13}^* & \sigma_{13}^* & \sigma_{03}^* \\
 \sigma_{01}^* & \sigma_{11} & \sigma_{11} & \sigma_{13} & \sigma_{11} & \sigma_{15} & \sigma_{16} & \sigma_{11} & \sigma_{11} & \sigma_{16} & \sigma_{15} & \sigma_{11} & \sigma_{13} & \sigma_{11} & \sigma_{11} & \sigma_{01}^* \\
 \sigma_{01}^* & \sigma_{11} & \sigma_{11} & \sigma_{13} & \sigma_{11} & \sigma_{15} & \sigma_{16} & \sigma_{11} & \sigma_{11} & \sigma_{16} & \sigma_{15} & \sigma_{11} & \sigma_{13} & \sigma_{11} & \sigma_{11} & \sigma_{01}^* \\
 \sigma_{00} & \sigma_{01} & \sigma_{01} & \sigma_{03} & \sigma_{01} & \sigma_{05} & \sigma_{06} & \sigma_{01} & \sigma_{01} & \sigma_{06} & \sigma_{05} & \sigma_{01} & \sigma_{03} & \sigma_{01} & \sigma_{01} & \sigma_{00} \\
\end{array}
\right) \label{eq:sigma}
\end{align}
\vspace*{2pt}
\normalsize
\end{figure*}

\begin{figure*}[!t]
\small
\begin{align}
\begin{split}
r_4 = \rho_{05} &= \frac{3 \Gamma ^2-\sqrt{4 \Gamma ^2+1} \sinh \left(\frac{1}{2 T}\right) \sinh \left(\frac{\sqrt{4 \Gamma ^2+1}}{2 T}\right)+\Gamma ^2 \cosh \left(\frac{\sqrt{4 \Gamma ^2+1}}{T}\right)+\left(4 \Gamma ^2+1\right) \cosh \left(\frac{1}{2 T}\right) \cosh \left(\frac{\sqrt{4 \Gamma ^2+1}}{2 T}\right)+1}{4 \left(4 \Gamma ^2+1\right) \left(\cosh \left(\frac{\sqrt{4 \Gamma ^2+1}}{T}\right)+\cosh \left(\frac{1}{T}\right)\right)}\\
r_8 = \rho_{15} &= -\frac{\Gamma  \sinh \left(\frac{\sqrt{4 \Gamma ^2+1}}{2 T}\right) \left(\sinh \left(\frac{\sqrt{4 \Gamma ^2+1}}{2 T}\right)+\sqrt{4 \Gamma ^2+1} \left(\cosh \left(\frac{\sqrt{4 \Gamma ^2+1}}{2 T}\right)+\sinh \left(\frac{1}{2 T}\right)+\cosh \left(\frac{1}{2 T}\right)\right)\right)}{4 \left(4 \Gamma ^2+1\right) \left(\cosh \left(\frac{\sqrt{4 \Gamma ^2+1}}{T}\right)+\cosh \left(\frac{1}{T}\right)\right)} \\
r_{13} = \rho_{55} &= \frac{\left(\frac{\sinh \left(\frac{\sqrt{4 \Gamma ^2+1}}{2 T}\right)}{\sqrt{4 \Gamma ^2+1}}+\cosh \left(\frac{\sqrt{4 \Gamma ^2+1}}{2 T}\right)+\sinh \left(\frac{1}{2 T}\right)+\cosh \left(\frac{1}{2 T}\right)\right)^2}{8 \left(\cosh \left(\frac{\sqrt{4 \Gamma ^2+1}}{T}\right)+\cosh \left(\frac{1}{T}\right)\right)}\\
s_4 = \sigma_{05} &= \frac{(\expval{\mathsf{ZZ}_{AB}}+2)^2 \left(4 \expval{\mathsf{XX}_{AB}}+\expval{\mathsf{ZZ}_{AB}}^2-4\right)}{64 \left(\expval{\mathsf{ZZ}_{AB}}^2+4\right)} \\
s_8 = \sigma_{15} &= \frac{(\expval{\mathsf{ZZ}_{AB}}+2) \left(\expval{\mathsf{X}_{A}}^2 \left(\expval{\mathsf{ZZ}_{AB}}^2+4\right)+4 \expval{\mathsf{ZZ}_{A}} \left(\expval{\mathsf{ZZ}_{AB}}^2-4\right)\right)}{64 \expval{\mathsf{X}_{A}} \left(\expval{\mathsf{ZZ}_{AB}}^2+4\right)} \\
s_{13} = \sigma_{55} &= \frac{(\expval{\mathsf{ZZ}_{AB}}+2)^2 \left(-8 \expval{\mathsf{ZZ}_{A}}+\expval{\mathsf{ZZ}_{AB}}^2+4\right)}{64 \left(\expval{\mathsf{ZZ}_{AB}}^2+4\right)}
\end{split}
\label{eq:pruned_elements}
\end{align}
\hrulefill
\vspace*{2pt}
\normalsize
\end{figure*}

\section{Performance of cost functions for varying $g$ \label{app:g_neq_1}}

\begin{figure*}[htbp!]
\includegraphics[width=\textwidth]{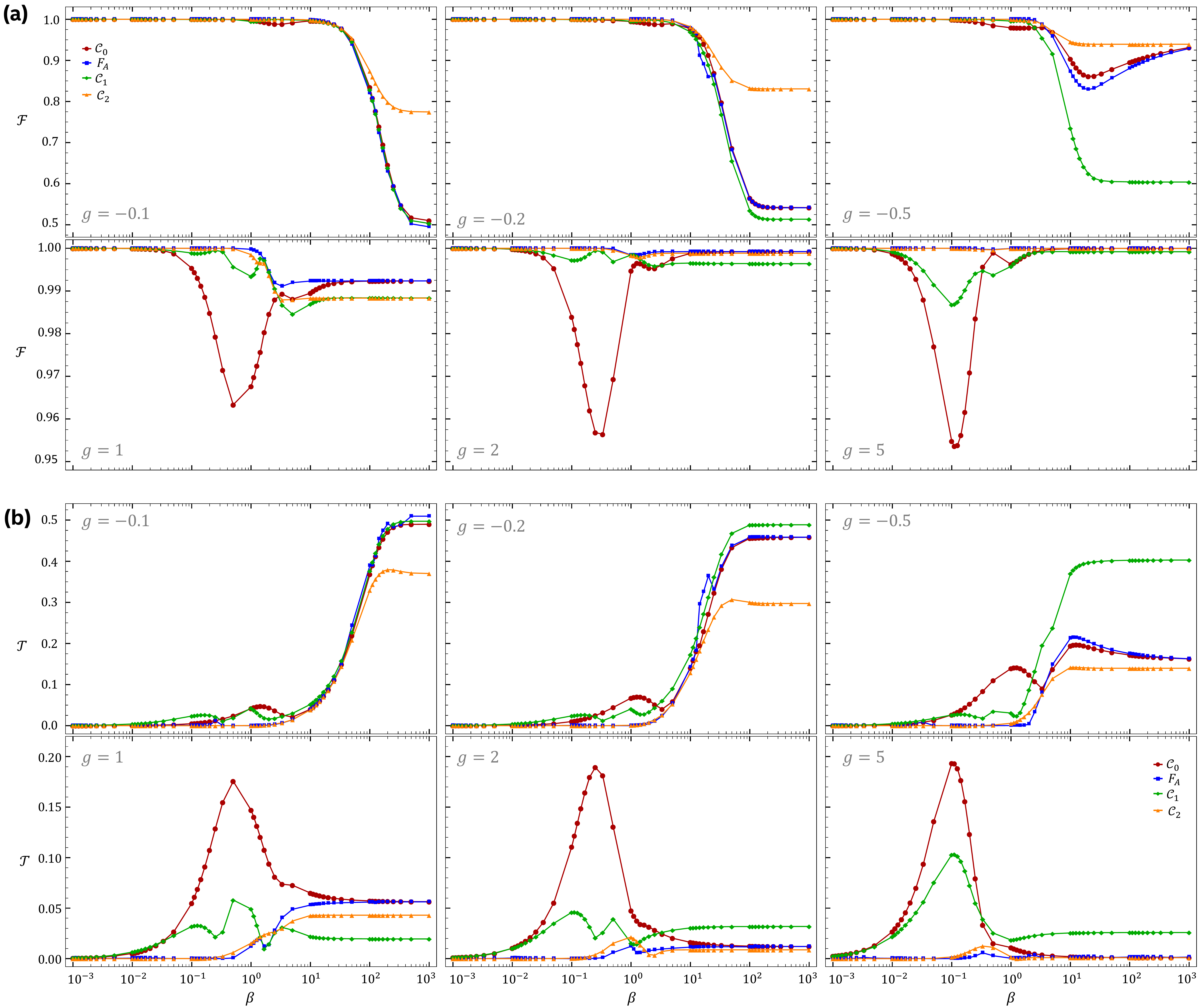}
\caption{Cost function performance comparison between $F_A$, $\mathcal{C}_0$, $\mathcal{C}_1$, and $\mathcal{C}_2$ using (a) fidelity $\mathcal{F}$ and (b) trace distance $\mathcal{T}$ as proximity measures for various transverse field strengths $g$.}
\label{fig:All_CF_vary_g}
\end{figure*}

In \cref{fig:All_CF_vary_g} we compare the different cost functions' performance at generating TFD states for different $g$ values. We find that $\mathcal{C}_2$ outperforms $F_A$ for most cases. Note that the low temperature performance of $\mathcal{C}_1$ is poor for the case of $g \neq 1$, indicating the optimal coefficients should be re-optimized for various $g$ values. Finding a general expression for the coefficients is desirable as the intermediate temperature performance is still superior to $\mathcal{C}_0$. We note that $F_A$ as a cost function occasionally had difficulty converging to a minimum, especially for $g=-0.1$ and $g=-0.2$, as can be seen from sudden jumps in the traces.

\section*{Acknowledgments}
\addcontentsline{toc}{section}{Acknowledgments}

The authors thank Sonika Johri and Xiang Chris Zou for insightful discussions.

\bibliographystyle{IEEEtrans}
\bibliography{bibliography}

\end{document}